\documentclass[twocolumn]{aastex631}

\usepackage{amssymb}
\usepackage{mathrsfs}
\usepackage{amsmath}

\begin{document}

\title{A Phase-resolved View of ``Heartbeat"-like variability in IGR J17091--3624 During the 2022 Outburst}

\correspondingauthor{Qing-Cang Shui}
\email{shuiqc@ihep.ac.cn}
\correspondingauthor{Shu Zhang}
\email{szhang@ihep.ac.cn}
\correspondingauthor{Jing-Qiang Peng}
\email{pengjq@ihep.ac.cn}

\author[0000-0001-5160-3344]{Qing-Cang Shui}
\affiliation{Key Laboratory of Particle Astrophysics, Institute of High Energy Physics, Chinese Academy of Sciences, 100049, Beijing, China}
\affiliation{University of Chinese Academy of Sciences, Chinese Academy of Sciences, 100049, Beijing, China}

%\author{Your Name}
%\affiliation{Key Laboratory of Particle Astrophysics, Institute of High Energy Physics, Chinese Academy of Sciences, 100049, Beijing, China}
%\affiliation{University of Chinese Academy of Sciences, Chinese Academy of Sciences, 100049, Beijing, China}
%\affiliation{Institut f\"{u}r Astronomie und Astrophysik, Kepler Center for Astro and Particle Physics, Eberhard Karls, Universit\"{a}t, Sand 1, D-72076 T\"{u}bingen, Germany}
\author{Shu Zhang}
\affiliation{Key Laboratory of Particle Astrophysics, Institute of High Energy Physics, Chinese Academy of Sciences, 100049, Beijing, China}

\author[0000-0002-5554-1088]{Jing-Qiang Peng}
\affiliation{Key Laboratory of Particle Astrophysics, Institute of High Energy Physics, Chinese Academy of Sciences, 100049, Beijing, China}
\affiliation{University of Chinese Academy of Sciences, Chinese Academy of Sciences, 100049, Beijing, China}

\author[0000-0001-5586-1017]{Shuang-Nan Zhang}
\affiliation{Key Laboratory of Particle Astrophysics, Institute of High Energy Physics, Chinese Academy of Sciences, 100049, Beijing, China}
\affiliation{University of Chinese Academy of Sciences, Chinese Academy of Sciences, 100049, Beijing, China}

\author[0000-0001-8768-3294]{Yu-Peng Chen}
\affiliation{Key Laboratory of Particle Astrophysics, Institute of High Energy Physics, Chinese Academy of Sciences, 100049, Beijing, China}

\author[0000-0003-3188-9079]{Ling-Da Kong}
\affiliation{Institut f\"{u}r Astronomie und Astrophysik, Kepler Center for Astro and Particle Physics, Eberhard Karls, Universit\"{a}t, Sand 1, D-72076 T\"{u}bingen, Germany}

\author{Zhuo-Li Yu}
\affiliation{Key Laboratory of Particle Astrophysics, Institute of High Energy Physics, Chinese Academy of Sciences, 100049, Beijing, China}

\author[0000-0001-9599-7285]{Long Ji}
\affiliation{School of Physics and Astronomy, Sun Yat-Sen University, Zhuhai, 519082, China}

\author[0000-0002-6454-9540]{Peng-Ju Wang}
\affiliation{Institut f\"{u}r Astronomie und Astrophysik, Kepler Center for Astro and Particle Physics, Eberhard Karls, Universit\"{a}t, Sand 1, D-72076 T\"{u}bingen, Germany}

\author[0000-0003-4856-2275]{Zhi Chang}
\affiliation{Key Laboratory of Particle Astrophysics, Institute of High Energy Physics, Chinese Academy of Sciences, 100049, Beijing, China}

\author[0000-0002-0638-088X]{Hong-Xing Yin}
\affiliation{Shandong Key Laboratory of Optical Astronomy and Solar-Terrestrial Environment, School of Space Science and Physics, Institute of Space Sciences, Shandong University, Weihai, Shandong 264209, China}

\author{Jian Li}
\affiliation{CAS Key Laboratory for Research in Galaxies and Cosmology, Department of Astronomy, University of Science and Technology of China, Hefei 230026, China}
\affiliation{School of Astronomy and Space Science, University of Science and Technology of China, Hefei 230026, China}

%\affiliation{Yunnan Observatories, Chinese Academy of Sciences, Kunming 650216, P.R. China}


\begin{abstract}
IGR J17091--3624, in addition to GRS 1915+105, is the only black hole X-ray binary that displays ``heartbeat"-like variability characterized by structured flares with high amplitudes. In this study, we conduct a detailed phase-resolved analysis of the recently identified ``heartbeat"-like Class \uppercase\expandafter{\romannumeral10} variability in IGR J17091--3624 during its 2022 outburst, utilizing data from NICER and NuSTAR observations. A shortage in the high-energy ($>$20 keV) X-ray flux is detected at peak phases of the soft X-ray flare at a $\sim15\sigma$ confidence level from the phase-folded light curves. Furthermore, our phase-resolved spectral analysis reveals variations in the spectral shape, particularly showing significant synchronous variations in the disk temperature and flux with the count rate. These findings imply that the flare is primarily driven by instabilities within the accretion disk, consistent with previous studies on the well-known Class $\rho$ variability in GRS 1915+105. However, we also observe a positive correlation between the disk temperature and flux over the flare cycle, which differs from a loop relation between the two parameters found in the Class $\rho$ variability. This could suggest differences in underlying physical processes between the two variability classes. Variations in the Componization component during flares are also observed: the electron temperature and covering fraction show anti-correlations with the disk flux, revealing potential interactions between the accretion disk and the corona during these flares.
\end{abstract}

\keywords{Accretion (14) --- Black hole physics (159) --- X-ray binary stars (1811) --- Stellar mass black holes (1611)}


\section{Introduction} \label{sec:intro}
Black hole X-ray binaries (BHXRBs) can display variability across multiple time scales. In the case of typical BHXRBs, the primary feature is undergoing recurrent outbursts lasting from months to years, stemming from thermal-viscous instabilities in the accretion disk \citep{2001A&A...373..251D}. Throughout an outburst, the source exhibits transitions in X-ray spectral and timing properties \citep{2006ARA&A..44...49R,2007A&ARv..15....1D}. These outbursts can be canonically categorized into distinct states, each characterized by unique X-ray spectral and timing properties \citep[][]{2005A&A...440..207B,2005Ap&SS.300..107H}. In the hard state, the emitted spectrum is predominantly characterized by a nonthermal power-law component, whereas in the soft state, a dominant thermal multi-temperature blackbody component emerges \citep{2006ARA&A..44...49R}. The multi-temperature blackbody component is thought to originate from a geometrically thin and optically thick accretion disk \citep{1973A&A....24..337S,1974MNRAS.168..603L}. The nonthermal power-law component, on the other hand, arises from the Comptonization of soft photons within an extended cloud of hot electrons, commonly known as the X-ray corona \citep{1975ApJ...195L.101T,1980A&A....86..121S,1994ApJ...434..570T,1996MNRAS.283..193Z}, and/or in a jet base \citep{2005ApJ...635.1203M,2021NatCo..12.1025Y}. However, the jet models encounter challenges associated with energetics and recent constraints on polarization \citep[see e.g.][]{2005MNRAS.360L..68M,2009MNRAS.400.1512M,2022Sci...378..650K,2024ApJ...968...76I}. On shorter time scales, low frequency quasi-periodic oscillations \citep[LFQPOs, roughly 0.1--30 Hz,][]{1989ARA&A..27..517V} have been consistently observed in the majority of BHXRBs. These characteristics are defined by a narrow peak with finite width in the power spectral density (PSD), and are classified as types A, B, and C based on centroid frequency, quality factor, and root-mean-square (rms) amplitude. \citep{1999ApJ...526L..33W,2005ApJ...629..403C,2006ARA&A..44...49R}.

IGR J17091--3624 and GRS 1915+105 are distinguished from other canonical X-ray binaries by their unique X-ray variability patterns, characterized by highly structured flares and dips with high amplitudes \citep{1996ApJ...473L.107G,2011ApJ...742L..17A}. These variability patterns observed in the X-ray light curves are commonly categorized into various distinct classes: 9 in IGR J17091--3624 \citep{2017MNRAS.468.4748C}, and 14 in GRS 1915+105 \citep{2000A&A...355..271B,2002MNRAS.331..745K,2005A&A...435..995H}. The most interesting and famous among the various variability classes is the ``heartbeat" variability (Class \uppercase\expandafter{\romannumeral4} in IGR J17091--3624 and Class $\rho$ in GRS 1915+105), where the X-ray light curve matches with the electrocardiogram of a human being. In this study, we will refer to structured variabilities with high amplitude as ``heartbeat"-like. The ``heartbeat"-like variabilities have commonly been interpreted as limit-cycle instabilities within the inner accretion disk, likely stemming from the radiation pressure instability \citep{1998MNRAS.298..888S,2000ApJ...542L..33J,2000ApJ...535..798N,2004MNRAS.349..393D}.

IGR J17091--3624, a transient Galactic black hole candidate discovered by INTEGRAL in 2003 \citep{2003ATel..149....1K}, exhibited a diverse range of ``heartbeat"-like variability behaviors during the bright outburst in 2011, with the detection of both low- and high-frequency QPOs \citep{2011ApJ...742L..17A,2011A&A...533L...4R,2012ApJ...747L...4A}. If the ``heartbeat"-like variability originates from the radiation pressure instability, theoretical models predict that the source is accreting at a high Eddington ratio \citep[e.g. $\gtrsim26\%\dot{M}_{\rm Edd}$,][]{2000ApJ...535..798N}. Considering that the observed flux from the source was much lower than GRS 1915+105, this implies the lowest-mass black hole known ($M_{\rm BH}<3M_{\odot}$ if $d<17$ kpc) in IGR J17091--3624, or it is very distant, possibly even outside the stellar disk of our galaxy. In February 2016, IGR J17091--3624 experienced another outburst \citep{2016ATel.8742....1M}. By modeling the reflection spectra in the hard state using NuSTAR and Swfit/XRT data, \citet{2017ApJ...851..103X} reported an inclination angle of the accretion disk of $\sim30-40^\circ$, a result consistent with the inclination of $\sim45^\circ$ obtained by \citet{2018MNRAS.478.4837W} based on a more extensive dataset. Additionally, \citet{2018MNRAS.478.4837W} identified a ``heartbeat"-like variability pattern in one NuSTAR observation, and subsequently conducted phase-resolved spectral analysis, revealing modulations in the reflected emission. During the more recent outburst in 2022 \citep{2022ATel15282....1M}, Neutron Star Interior Composition Interior Explorer \citep[NICER,][]{2016SPIE.9905E..1HG} monitored the rapid state transition of IGR J17091--3624 \citep{2022ATel15287....1W,2024ApJ...963...14W}. Utilizing NICER and Nuclear Spectroscopic Telescope Array \citep[NuSTAR,][]{2013ApJ...770..103H} observations, two types of ``heartbeat"-like variability (Class \uppercase\expandafter{\romannumeral5} and new Class \uppercase\expandafter{\romannumeral10}) were identified by \citet{2024ApJ...963...14W}. In this study, we conduct a comprehensive phase-resolved analysis of the new Class \uppercase\expandafter{\romannumeral10} ``heartbeat"-like variability with NICER and NuSTAR observations. We provide an overview of the observations and our data reduction in Section~\ref{sec:2}, followed by the presentation of the timing and spectral analyses in Section~\ref{sec:3}. Finally, we discuss and summarize these results in Section~\ref{sec:4} and Section~\ref{sec:5}, respectively.

\section{Observations and Data Reduction} \label{sec:2}
\subsection{Selection of the ``heartbeat" state data sets}
\begin{figure}
\centering
    \includegraphics[width=\linewidth]{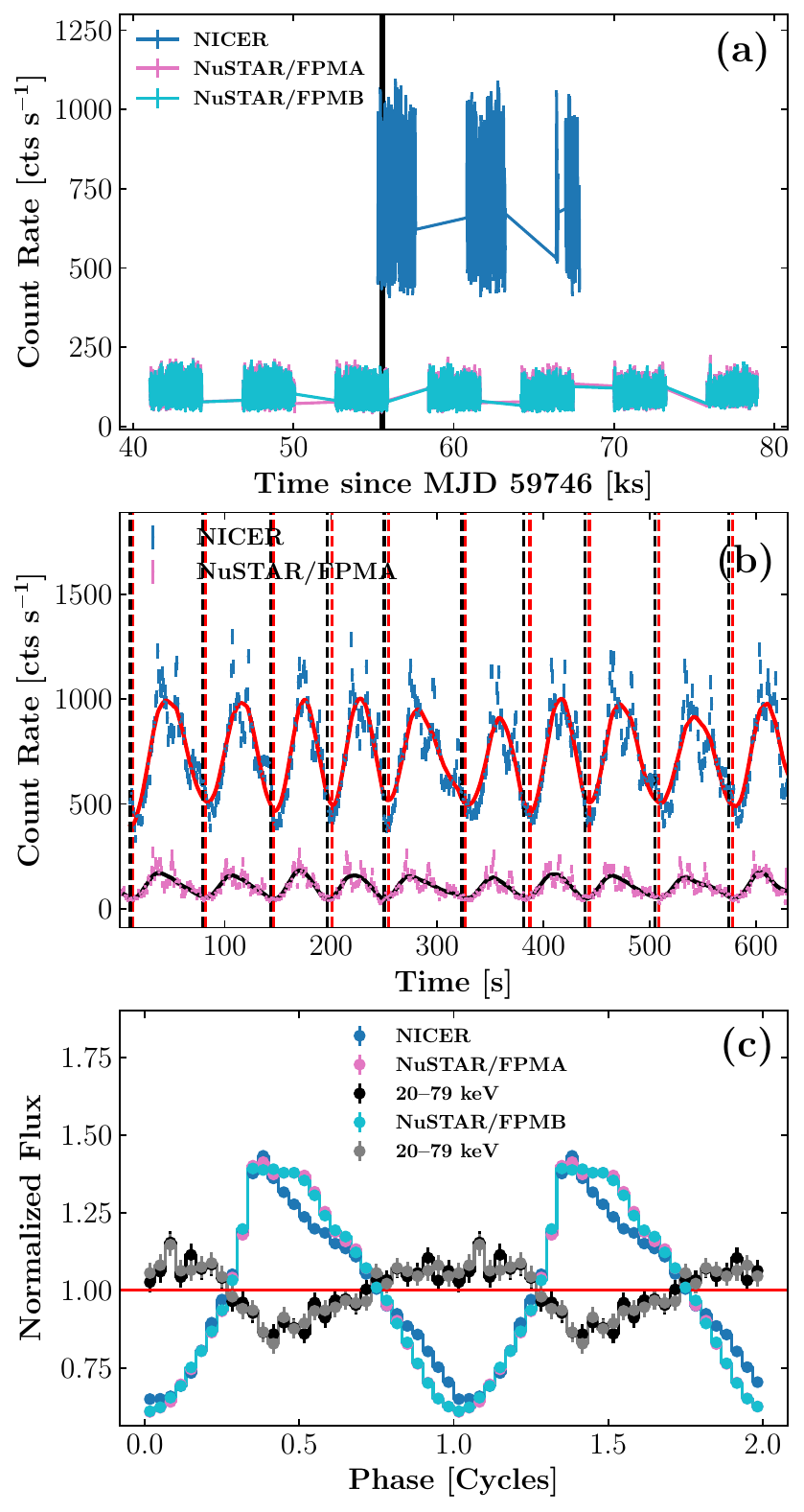}
    \caption{Results of the timing analysis. (a) Light curves of IGR J17091--3624 with a 10-s time resolution from the quasi-simultaneous observations of NICER (1--10 keV), NuSTAR/FPMA (3--79 keV) and FPMB (3--79 keV). (b) Representative $\sim700$ s light curves with a 1-s time bin from the simultaneous observations of NICER (1--10 keV) and NuSTAR/FPMA (3--79 keV), and the low-pass-filtered light curves (colored solid lines) obtained from HHT analysis. The vertical red and black dashed lines represent minima identified by HHT analyses of NICER and NuSTAR data respectively. (c) HHT phase-folded light curves of NICER (in the 1--10 keV energy band) and NuSTAR (in the 3--79 keV and 20--79 keV energy bands) data with 30 phase bins per cycle. }
    \label{fig:1}
\end{figure}

IGR J17091--3624 underwent a new outburst in March 2022 \citep{2022ATel15282....1M}. During this outburst, NICER conducted near-daily observations of the source from March 27 to August 21, 2022 \citep{2024ApJ...963...14W,2024ApJ...963..118W}. Analysis of all 136 NICER observations by \citet{2024ApJ...963...14W} identified two classes of ``heartbeat"-like variability, Class \uppercase\expandafter{\romannumeral5} and a newly identified Class \uppercase\expandafter{\romannumeral10}. Class \uppercase\expandafter{\romannumeral5} variability was previously identified in the 2011 outburst, characterized by repeated, sharp, high-amplitude flares \citep{2011ApJ...742L..17A}. However, the recurrence time of these flares exhibits significant drift even within a 500 s segment \citep[see][]{2024ApJ...963...14W}. Furthermore, Class V flares can be categorized into two distinct populations, similar to the behavior of Class $\mu$ in GRS 1915+105 \citep{2017MNRAS.468.4748C}. In contrast, the new pattern of variability, Class \uppercase\expandafter{\romannumeral10}, exhibits nearly sinusoidal modulation with large amplitude, uniformity, and symmetric flares \citep[see][for details]{2024ApJ...963...14W}. Therefore, this study will focus on conducting phase-resolved analyses for Class \uppercase\expandafter{\romannumeral10} variability in a more direct and efficient manner. As reported by \citet{2024ApJ...963...14W}, Class \uppercase\expandafter{\romannumeral10} variability was observed in NICER observations from June 15 to July 18, 2022. We identify 15 observations showing this new ``heartbeat"-like variability in a quick-look analysis of the NICER light curves (refer to Table~\ref{tab:1} for details of the NICER observations used in this study). Here, we focus specifically on Class \uppercase\expandafter{\romannumeral10} variability and conduct a thorough timing and spectral analysis of these observations. On June 16, 2022, NuSTAR carried out a quasi-simultaneous observation with NICER, capturing the Class \uppercase\expandafter{\romannumeral10} variability as well. Therefore, this NuSTAR observation is included in our analysis. Figure~\ref{fig:1} (a) illustrates the light curves of the NICER and NuSTAR observations on June 16, 2022 with a 10-second time resolution.

\begin{table}[]
    \centering
    \caption{Log of NICER Observations of IGR J17091--3624 Used in This Work.\label{tab:1}}
    \begin{tabular}{cccc}
    \hline
    \hline
    \# & Observation ID & Start Time & Exposure Time\\ 
       &                & (MJD) & (s)\\
    \hline
    1 & 5618010802 & 59745.09 & 5582 \\
    2 & 5618010803 & 59746.64 & 5746 \\
    3 & 5618010804 & 59747.75 & 2401 \\
    4 & 5618010805 & 59748.45 & 3823 \\
    5 & 5618011005 & 59763.09 & 980 \\
    6 & 5618011006 & 59764.06 & 1927 \\
    7 & 5618011007 & 59765.02 & 789 \\
    8 & 5618011101 & 59766.06 & 2069 \\
    9 & 5618011102 & 59767.67 & 698  \\
    10 & 5618011103 & 59768.39 & 656  \\
    11 & 5618011201 & 59774.58 & 1098 \\
    12 & 5618011202 & 59775.16 & 795  \\
    13 & 5618011203 & 59776.25 & 876  \\
    14 & 5618011204 & 59777.16 & 884  \\
    15 & 5618011205 & 59778.06 & 2854 \\
    \hline
    \hline
    \end{tabular}
\end{table}

\subsection{Data Reduction}
\subsubsection{NICER}
NICER, a payload on the International Space Station (ISS), is dedicated to timing and spectroscopic analysis at soft X-rays \citep[0.2--12 keV;][]{2016SPIE.9905E..1HG}. The data reduction is performed by using the official software \texttt{HEASOFT V6.31.1/NICERDAS} package (version 2022-12-16\_V010a) with the latest CALDB \texttt{xti20221001}. We first process the data with the standard pipeline processing tool \texttt{nicerl2}. Subsequently, the light curves and energy spectrum are extracted using \texttt{nicerl3-lc} and \texttt{nicerl3-spect} tools, respectively. The \texttt{nibackgen3C50} model \citep{2022AJ....163..130R} is used to estimated the background corresponding to the observation. Here, Focal Plane Modules (FPM) 14 and 34 are always excluded, since they often exhibited episodes of increased detector noise. The adopted energy range in the spectral analysis is 1--10 keV, and a systematic error of 1 per cent is incorporated for 1--3 keV energy band to reduce the effects from calibration uncertainties at low energies.

\subsubsection{NuSTAR}
NuSTAR, the first mission to use focusing telescopes to image the sky in the high-energy X-ray (3--79 keV) region of the electromagnetic spectrum, was launched at 9 am PDT, June 13, 2012 \citep{2013ApJ...770..103H}. The data are processed using the NuSTAR Data Analysis Software (\texttt{NuSTARDAS v2.1.2}) with the latest calibration data base (\texttt{CALDB 20230613}). At first, the standard pipeline routine \texttt{nupipeline} is used for data reduction. Since IGR J17091--3624 was bright in the Class \uppercase\expandafter{\romannumeral10} state ($\gtrsim100\ \rm{cts\ s^{-1}}$), an additional status expression \texttt{STATUS == b0000xxx00xxxx000}\footnote{\url{https://heasarc.gsfc.nasa.gov/docs/nustar/analysis/}} is applied in this step. Then, \texttt{nuproducts} is used to generate light curves, spectra, response matrix, and ancillary response files, where source and background events are extracted from a 100 arcsec circular region and a 130 arcsec source-free circular region, respectively. Finally, we group the spectra via the \texttt{grppha} task to ensure a signal-to-noise ratio of at least five for each bin. In the spectral analysis, the fitted energy range is 3--50 keV, since the spectrum is soft in the Class \uppercase\expandafter{\romannumeral10} state, and therefore the spectrum above 50 keV is dominated by the background.  

\section{Analysis and Results}
\label{sec:3}
\subsection{Timing Analysis}
We first perform a timing analysis for the quasi-simultaneous observations of NICER and NuSTAR in the Class \uppercase\expandafter{\romannumeral10} state. In Figure~\ref{fig:1} (b), we present representative $\sim700$ s simultaneous light curves of NICER and NuSTAR observations with a 1-s time resolution as scatter points accompanied by error bars. The position of this time interval within the entire observational period is indicated by the solid black line plotted in Figure~\ref{fig:1} (a). Both NICER and NuSTAR light curves clearly exhibit strong, highly structured variability with large amplitude. The recurrence time and amplitude of the flare vary among adjacent ones, and sometimes additional mini-flares can occur at the peaks of major flares. However, the relative phases among major flares and mini-flares could be unstable. 

Since the recurrence time and amplitude of the flare are variable within 100-s time scale, folding on a period to construct the phase-resolved spectra is not suitable for this variability pattern. Considering that the feature of the new Class \uppercase\expandafter{\romannumeral10} variability is the uniformity of the flare time scale, leading to a narrow peak in the power spectra like QPOs \citep[see Figure 4 of][]{2024ApJ...963...14W}, we adopt the procedure outlined by \citet{2023ApJ...957...84S} that performing the Hilbert-Huang transform (HHT) analysis to construct its instantaneous phase function. The HHT method, initially introduced by \citet{1998RSPSA.454..903H} as an adaptive data analysis technique, provides a powerful tool for studying signals with non-stationary periodicity \citep{1998RSPSA.454..903H,2008RvGeo..46.2006H}. This approach comprises two main components: mode decomposition and Hilbert spectral analysis (HSA). The mode decomposition aims to decompose a time series into several intrinsic mode functions (IMFs), and the HSA allows obtaining both the frequency and phase functions for the desired IMFs, such as the LFQPO \citep[e.g.][]{2014ApJ...788...31H,2020ApJ...900..116H,2023ApJ...951..130Y,2023ApJ...957...84S,2024ApJ...965L...7S} and the ``heartbeat''-like variability. We conduct the HHT analysis on the 1-s time-bin light curves within the 1--10 keV energy range from NICER data and the 3--79 keV energy range from NuSTAR data. By utilizing the variational mode decomposition \citep[VMD,][]{6655981}, the intrinsic variability of the ``heartbeat'' can be identified as the second IMF. Based on these decomposed IMFs, we present in Figure~\ref{fig:1} (b) the low-pass-filtered light curves as solid red and black lines for NICER and NuSTAR/FPMA data respectively. Through application of the HSA, we derive instantaneous phase functions for the variability. The vertical red and black dashed lines represent minima identified by HHT analyses of NICER and NuSTAR/FPMA data respectively. It is evident that the obtained phase functions from our HHT analyses with NICER and NuSTAR data exhibit strong consistency.

With the acquired phase functions, the HHT phase-folding method is capable of producing folded light curves. As shown in Figure~\ref{fig:1} (c), the phase-folded light curves of NICER (1--10 keV) and NuSTAR (3--79 keV) are consistent, displaying a slightly non-sinusoidal nature, characterized by a \textit{fast rise and slow decay} feature. The fractional root-mean-square (rms) amplitude of the ``heartbeat''-like variability is $\sim25\%$, consistent with the result obtained from the PSD analysis \citep[see][]{2024ApJ...963...14W}. We also present phase-folded light curves in 20--79 keV energy range for the NuSTAR data. Interestingly, both the high-energy folded light curves from FPMA and FPMB data show a rough anti-correlation with the low-energy ones. To be more precise, during the trough phases ($\sim0.75-1.25$ cycles) of the soft X-ray flare, the high-energy folded light curves remains relatively constant, while during the peak phases ($\sim0.25-0.75$ cycles), the high-energy count rate reduces to 75\% of that during the flare trough phases. The significance of the deficits in hard X-ray count rates of FPMA and FPMB in 20--79 keV are respectively estimated to be $13.5\sigma$ and $14.4\sigma$ using $(N_{\rm T}-N_{\rm P})/\sqrt{N_{\rm T}+N_{\rm P}}$, where $N_{\rm T}$ is the total counts in 20--79 keV during the trough phases ($\sim0.75-1.25$ cycles), and $N_{\rm P}$ is that during the peak phases ($\sim0.25-0.75$ cycles).

\begin{figure*}
\centering
\begin{minipage}[c]{0.48\linewidth}
\centering
    \includegraphics[width=\linewidth]{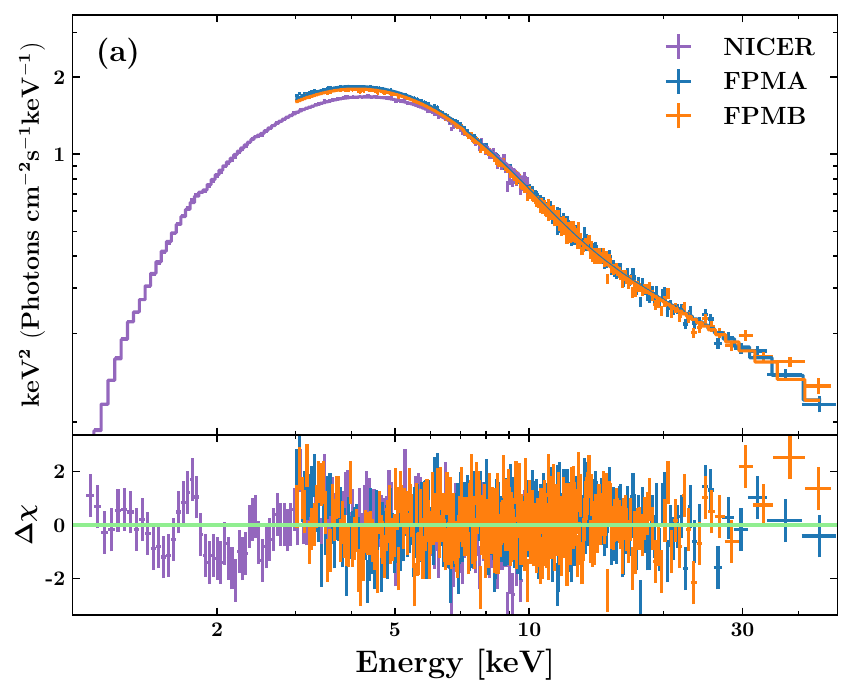}
\end{minipage}
\begin{minipage}[c]{0.48\linewidth}
\centering
    \includegraphics[width=\linewidth]{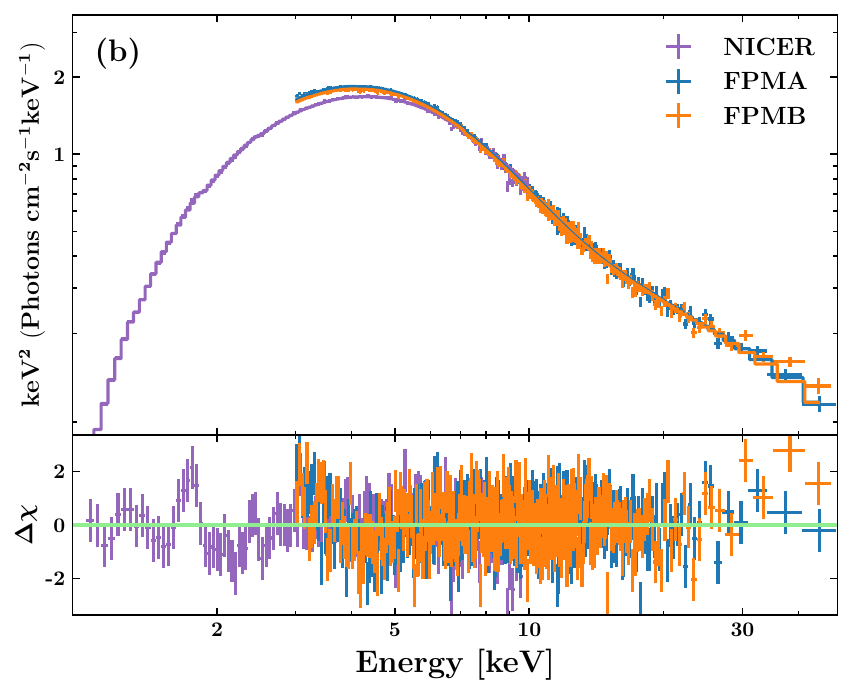}
\end{minipage}\\
\centering
\begin{minipage}[c]{0.48\linewidth}
\centering
    \includegraphics[width=\linewidth]{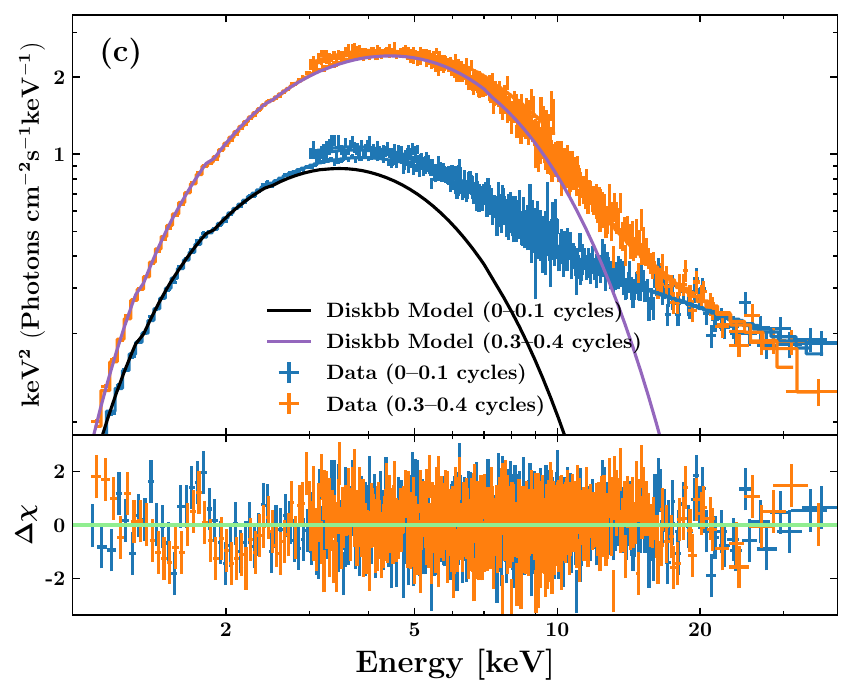}
\end{minipage}
\begin{minipage}[c]{0.48\linewidth}
\centering
    \includegraphics[width=\linewidth]{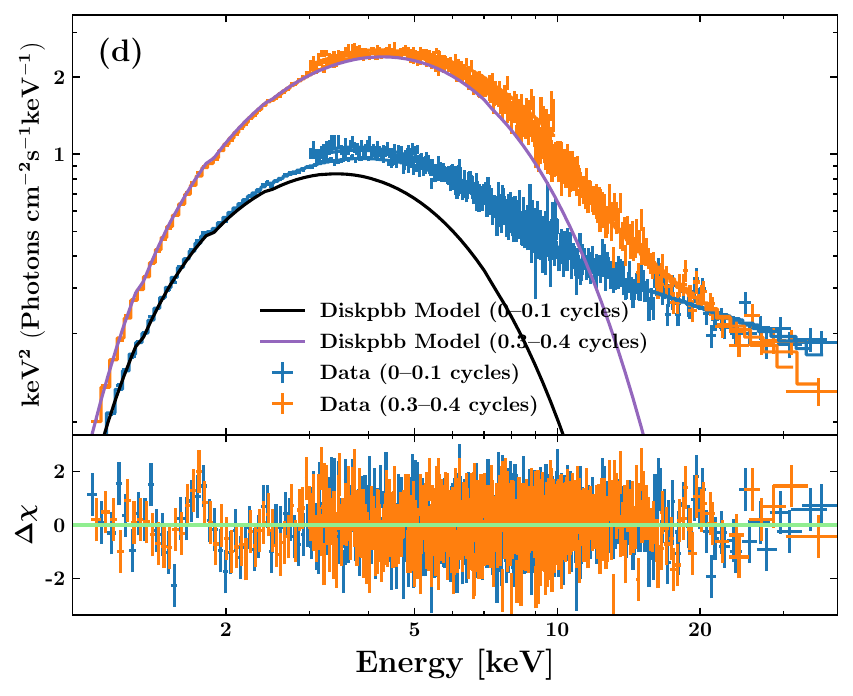}
\end{minipage}
    \caption{Spectral fittings (top panels) and corresponding residual plots (bottom panels). Panels (a) and (b) display phase-averaged spectral analyses using Models 1 and 2, respectively. Panels (c) and (d) show phase-resolved spectral analyses with Models 1 and 2, respectively. We specifically illustrate the peak phases and trough phases of the flare as representative examples. To enhance the demonstration of the phase dependence of the spectral component, we include solid lines representing the best-fitted \texttt{Diskbb} and \texttt{Diskpbb} models in panels (c) and (d).} \label{fig:2}
\end{figure*} 

\subsection{Energy Spectral Analysis}
The spectral analysis is performed using \texttt{XPSEC v12.14.0} software package \citep{1996ASPC..101...17A}. We first fit the joint NICER and NuSTAR (FPMA and FPMB) spectra with an absorbed convolution thermal-Comptonization model, incorporating input seed photons contributed by the spectral component \texttt{Diskbb}, which is available as \texttt{Thcomp} (a more accurate version of \texttt{Nthcomp}) in \texttt{XPSEC} \citep{2020MNRAS.492.5234Z}. This thermal-Comptonization model is described by the photon index ($\Gamma$), electron temperature ($kT_{\rm e}$) and covering fraction ($f_{\rm c}$). The galactic absorption is accounted for by the \texttt{Tbabs} model. In addition, we employ \texttt{Crabcorr} model to facilitate cross-calibration among NICER, NuSTAR spectra \citep{2010ApJ...718L.117S,2020ApJ...899...44W,2022ApJ...928...11Z,2024ApJ...960L..17P,2024MNRAS.527.8029Y}. This model introduces corrections to both the slope of the power-law through the parameter $\Delta\Gamma$ and normalization, $N$, achieved by applying a power-law multiplication to each model component. By fitting with Model 1: \texttt{Tbabs*Crabcorr*Thcomp*Diskbb}, we obtain a $\chi^2/{\rm d.o.f}$ of 1268.97/1084. However, Figure~\ref{fig:2} (a) shows that the residuals are still strongly correlated below 3 keV. Specifically, it is evident that the correlated residual features are a edge-like shape near $\sim$2.2 keV and a Gaussian-like emission around 1.8 keV. These energies correspond to specific features in NICER's effective area versus energy, with the 1.8 keV and 2.2 keV features attributed to silicon and gold, respectively \citep[see also][]{2020ApJ...899...44W}. Therefore, we posit that these remaining features stem from systematic effects within NICER's calibration. In order to investigate the possible deviation from the standard hypothesis, we implement the \texttt{Diskpbb} model \citep{1994ApJ...426..308M} as an alternative to the traditional \texttt{Diskbb}. This model introduces an exponent, $p$, for the radial temperature dependence ($T(r)\propto r^{-p}$), along with $T_{\rm in}$ and normalization as fit parameters.  In the context of a standard Shakura-Sunyaev disk, $p$ is typically around 0.75. However, various factors such as general relativistic effects, electron scattering, and advection can lead to different values of $p$. For instance, in the case of a slim disk, $p$ is approximately 0.5 \citep{2000PASJ...52..133W}. By employing Model 2: \texttt{Tbabs*Crabcorr*Thcomp*Diskpbb} for spectral fitting, we also obtain a reasonable $\chi^2/{\rm d.o.f}$ value of 1262.40/1083. The best-fitting value for $p$ is found to be $0.77\pm0.01$, which closely aligns with the standard hypothesis of the Shakura-Sunyaev disk model.

\begin{figure*}
\centering
\begin{minipage}[c]{0.48\linewidth}
\centering
    \includegraphics[width=\linewidth]{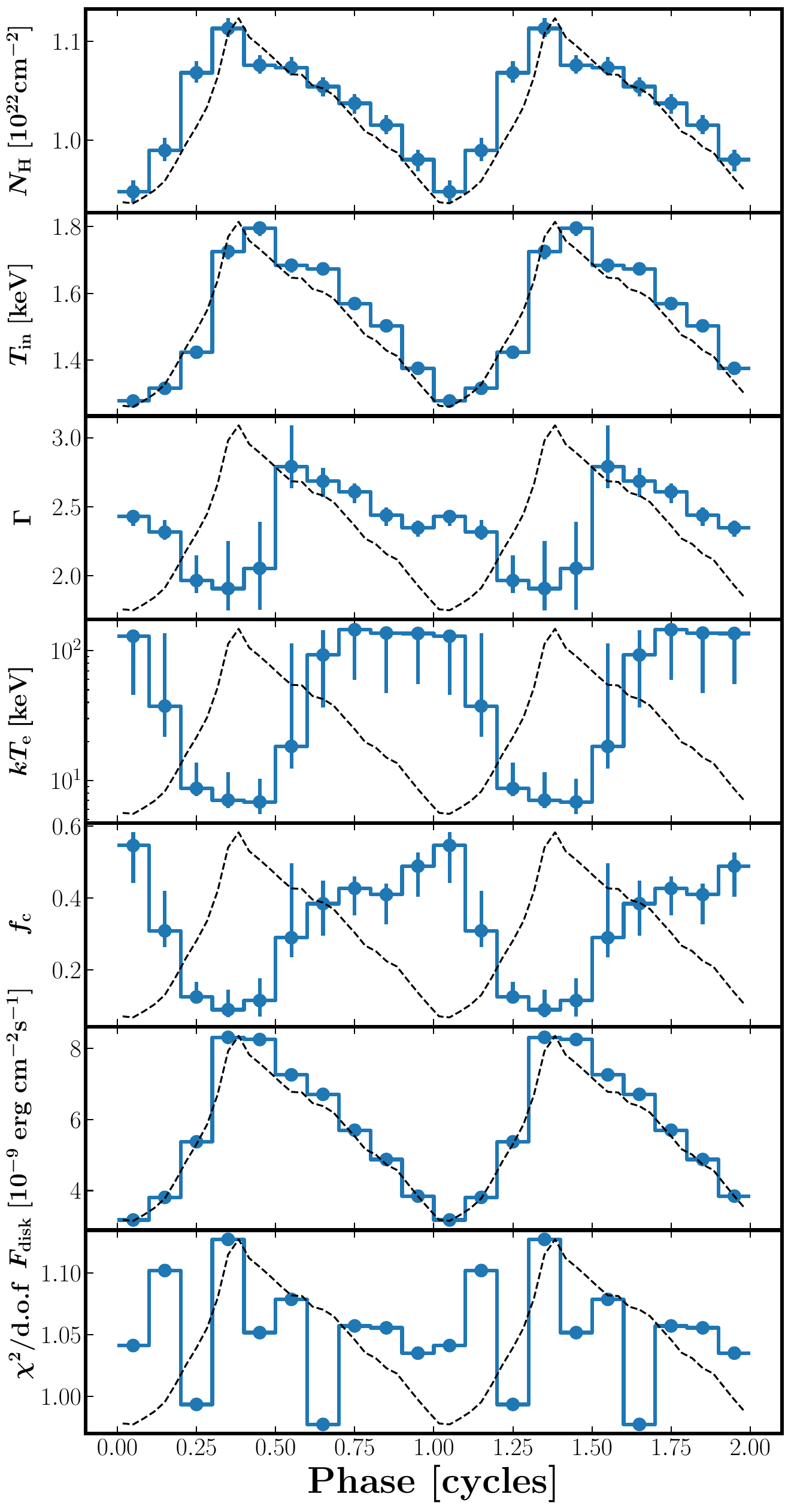}
\end{minipage}
\begin{minipage}[c]{0.48\linewidth}
\centering
    \includegraphics[width=\linewidth]{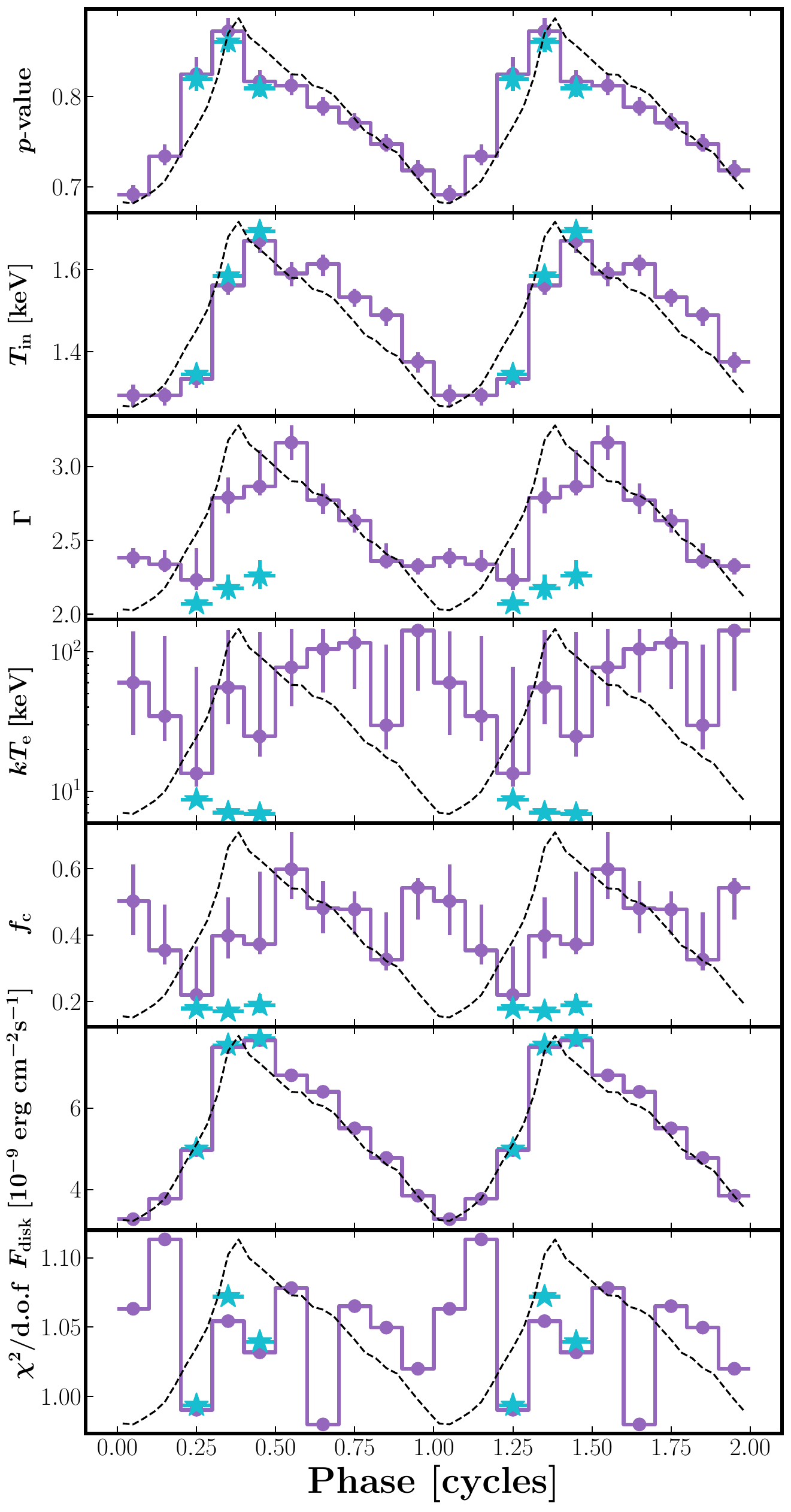}
\end{minipage}
    \caption{Variability of spectral parameters during the Class \uppercase\expandafter{\romannumeral10} variability cycle obtained from the fitting of joint NICER and NuSTAR phase-resolved spectra with Models 1 (left) and 2 (right). Additionally, we also include the phase-folded light curve from NICER data in each panel as the black dashed line. In the right panels, the cyan stars represent the corrected fitting results obtained by fixing the electron temperature ($kT_{\rm e}$) at the best-fitting values from Model 1 in the corresponding phase bins.} \label{fig:3}
\end{figure*} 

With the well-defined phase functions provided by the HHT analysis, we derive new GTIs corresponding to ten distinct phase intervals for NICER and NuSTAR observations. These updated GTIs allow us to extract spectra for each phase bin in order to conduct subsequent phase-resolved spectral analysis. In the phase-resolved spectral fitting using Model 1, we initially fix the equivalent hydrogen column density ($N_{\rm H}$) to the best-fit value obtained from the phase-averaged spectral fitting. However, we observe prominent residual features at low energies in certain phases. As a result, we treat $N_{\rm H}$ as a free parameter in the phase-resolved spectral fitting. This adjustment leads to reasonable reduced $\chi^2$ values. Furthermore, it is clear that the features below 3 keV have no significant phase dependence (see Figure~\ref{fig:2} (c)). We also conduct phase-resolved spectral fitting using Model 2, with $p$ as a free parameter. In this scenario, we achieve satisfactory goodness of fit despite fixing $N_{\rm H}$ at the best-fitting value obtained from the phase-averaged fitting (see Figure~\ref{fig:2} (d)). Additionally, we utilize the \texttt{cflux} model in \texttt{XSPEC} to compute the unabsorbed flux contribution from \texttt{Diskbb} and \texttt{Diskpbb} in Models 1 and 2. To compute the uncertainties, we utilize the Markov Chain Monte Carlo (MCMC) method using the Goodman-Weare algorithm with 32 walkers and a total length of 50,000 \citep{2010CAMCS...5...65G}, and the initial 2000 elements are discarded as a burn-in period. To further assess the convergence of the MCMC chain, we compare the one- and two-dimensional projections of the posterior distributions for each parameter from the first and second halves of the chain, and find no significant differences. In Appendix~\ref{appendix1}, we provide the contour maps and probability distributions for each free parameter in the phase-resolved spectral analysis with Models 1 and 2, respectively. We find that the electron temperature ($kT_{\rm e}$) is effectively constrained at the peak phases of the flare using Model 1, while spectral fitting with Model 2 does not yield effective constraints. However, neither of the two spectral modelings can effectively constrain $kT_{\rm e}$ at the flare trough phases.

\begin{figure}
\centering
    \includegraphics[width=\linewidth]{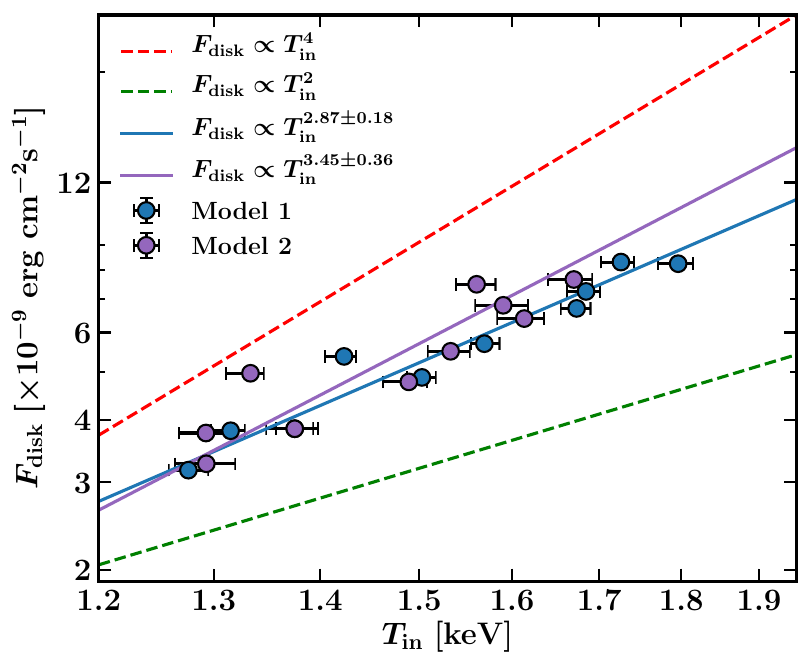}
    \caption{The disk flux $F_{\rm disk}$ vs. disk temperature $T_{\rm in}$ over the flare cycle from the fitting of the joint NICER and NuSTAR phase-resolved spectra with Models 1 (blue) and 2 (purple). The dashed lines represent the relations of $F_{\rm disk}\propto T^4_{\rm in}$ (red) and $F_{\rm disk}\propto T^2_{\rm in}$ (green), which are predicted by the standard Shakura-Sunyaev disk model and slim disk model, respectively. The solid lines represent the best-fitted $F_{\rm disk}\propto T^\alpha_{\rm in}$ models.} \label{fig:4}
\end{figure} 

Figure~\ref{fig:3} illustrates the spectral parameter variability throughout the ``heartbeat''-like variability cycle, as derived from fitting the joint NICER and NuSTAR phase-resolved spectra with Models 1 (left panels) and 2 (right panels). Additionally, the phase-folded light curve of NICER data is included in each panel as a black dashed line. The bottom two panels plot the reduced-$\chi^2$ from the phase-resolved spectral fitting for each phase bin. Both phase-resolved spectral analyses clearly show that $T_{\rm in}$ and disk flux ($F_{\rm disk}$) exhibit significant variations during the flare. Additionally, both results indicate that $T_{\rm in}$ and $F_{\rm disk}$ undergo synchronous modulations with the count rate, suggesting that the Class \uppercase\expandafter{\romannumeral10} ``heartbeat''-like variability is predominantly driven by the disk component. In Model 1 spectral fitting, we use the \texttt{Diskbb} model, which sets the radial temperature dependence to be $T(r)\propto r^{-3/4}$, while allowing $N_{\rm H}$ to vary as a free parameter. In this case, we observe a synchronous variation of $N_{\rm H}$ with the flux. However, when using Model 2 for spectral fitting, where we fix $N_{\rm H}$ and test the variability of radial temperature dependence, we find that the $p$ value also exhibits synchronous variation with the flux. Regarding the Comptonsized emission, discrepancies are observed in the results obtained from spectral fittings using two different models. While both sets of results suggest that the spectral index undergoes variations throughout the cycle, there are slight divergences in the patterns of $\Gamma$ modulations between the two results. The fitting with Model 1 shows that $\Gamma$ slightly decreases from $\sim2.5$ to $\sim1.9$ during the rising phases of the flare (0--0.5 cycles), followed by a sharp increase to $\sim2.8$ within a narrow phase interval ($\sim0.5$--0.6 cycles), and then gradually decreases to $\sim2.5$ at the end of flare. In contrast, when fitting with Model 2, $\Gamma$ remains relatively constant during the trough phases of the flare ($\sim0.75$--1.25 cycles), while it undergoes modulations within $\sim0.25$--0.75 cycles, reaching its maximum value at $\sim0.5$--0.6 cycles. In terms of the covering fraction ($f_{\rm c}$) and electron temperature ($kT_{\rm e}$), the analysis using Model 1 reveals that both $f_{\rm c}$ and $kT_{\rm e}$ exhibit anti-correlations with the disk flux. Conversely, no significant modulations in these two parameters are discernible from the modeling with Model 2. Due to the inability of Model 2 to effectively constrain the electron temperature over the flare cycles, as opposed to Model 1 which demonstrates strong constraints on this parameter at the peak phases (see Appendix~\ref{appendix1}), we opt to fix $kT_{\rm e}$ in the fittings with Model 2 in the three phase bins near the flare peak at the best-fitting values obtained from the fitting with Model 1 in the corresponding phase bins. This adjustment results in similar modulation trends of parameters in the Comptonization model across different spectral modelings, while having negligible impact on both the disk component and goodness of fit (see Figure~\ref{fig:3}). Additionally, this also suggests a weak constraint on the Comptonization component during Class \uppercase\expandafter{\romannumeral10} state, indicating that modulations in its parameters may be somewhat dependent on specific modeling choices.

\begin{figure*}
\centering
    \includegraphics[width=0.95\textwidth]{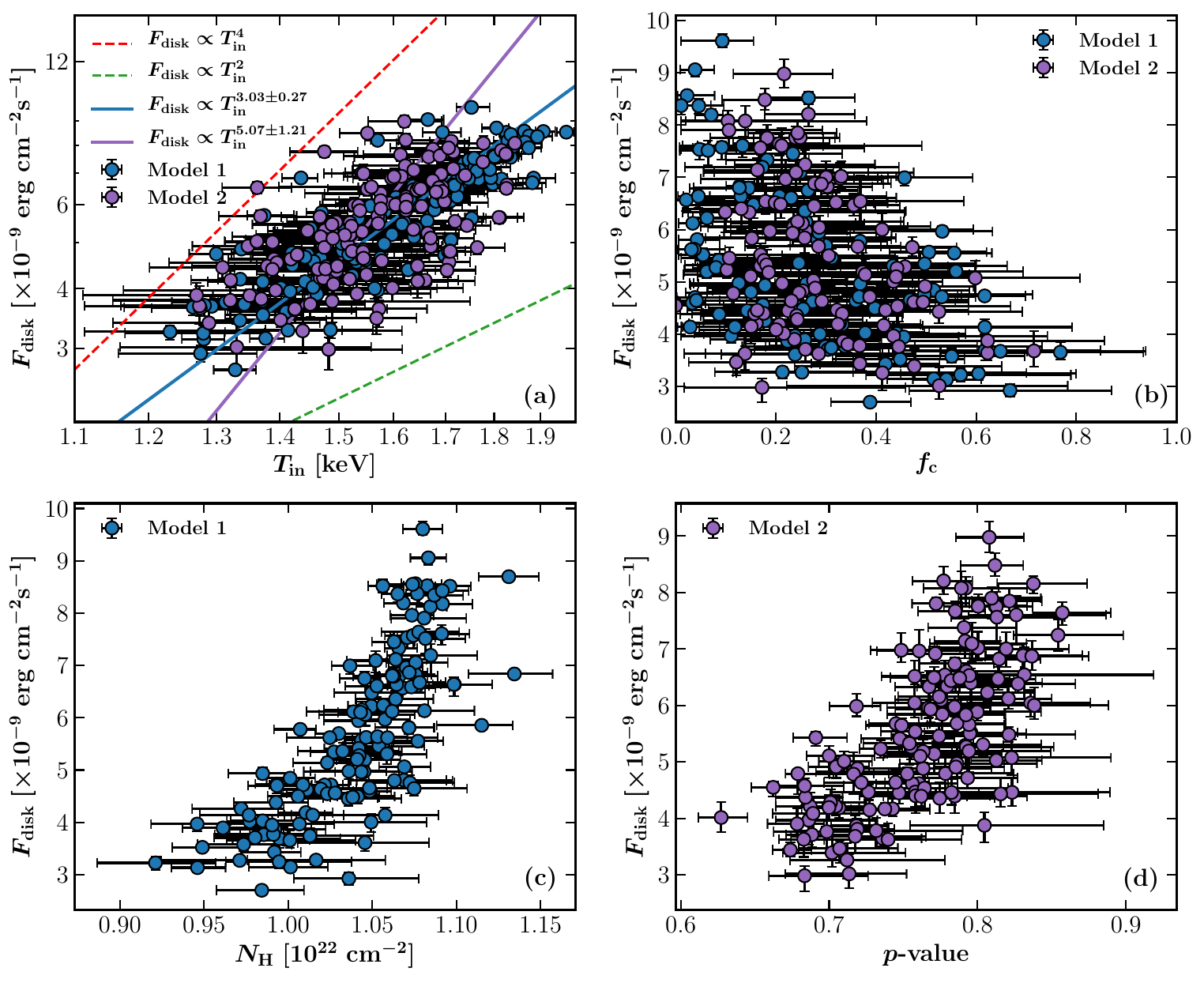}
    \caption{Plots of $F_{\rm disk}$ as a function of $T_{\rm in}$ (a), $f_{\rm c}$ (b), $N_{\rm H}$ (c) and $p$ (d) based on the fitting results obtained from all 150 NICER phase-resolved spectra with Model 1 (blue) and 2 (purple). In panel (a), the dashed lines represent the relations of $F_{\rm disk}\propto T^4_{\rm in}$ (red) and $F_{\rm disk}\propto T^2_{\rm in}$ (green), which are predicted by the standard Shakura-Sunyaev disk model and slim disk model, respectively. The solid lines represent the best-fitted $F_{\rm disk}\propto T^\alpha_{\rm in}$ models.} \label{fig:5}
\end{figure*}

As mentioned above, the phase-resolved spectral analyses reveal a prominent feature in the strong variability of the disk emission. Synchronous modulations with the count rate in both $T_{\rm in}$ and $F_{\rm disk}$ are observed, suggesting positive correlations between $T_{\rm in}$ and $F_{\rm disk}$ throughout the flare cycle. In Figure~\ref{fig:4}, we illustrate the behavior of the disk emission by plotting the disk flux against its temperature. We also conduct data fitting using the model $F_{\rm disk}\propto T_{\rm in}^\alpha$, where the power-law index $\alpha$ serves as an indicator of the nature of the accretion disk. In Figure~\ref{fig:4}, we also present theoretical $F_{\rm disk}\propto T_{\rm in}^4$ and $F_{\rm disk}\propto T_{\rm in}^2$ relationships, representing the predictions of standard Shakura-Sunyeav and slim disk models, respectively. The best-fitting values for $\alpha$ are $2.87\pm0.18$ and $3.45\pm0.36$ for Models 1 and 2, indicating that the disk emission obtained from spectral fitting with Model 1 significantly deviates from these two aforementioned theoretical models.

In Section~\ref{sec:2}, we have identified 15 NICER observations exhibiting the newly discovered ``heartbeat"-like variability through a preliminary analysis of light curves. To further validate the robustness of our findings from the phase-resolved analysis of joint NICER and NuSTAR data mentioned above, we conduct phase-resolved analyses on all 15 NICER observations and fit the resulting phase-resolved spectra with both Models 1 and 2. Each observation is divided into 10 phase bins based on the HHT analysis, resulting in a total of 150 phase-resolved spectra. For each observation, we perform a joint fitting for the 10 phase-resolved spectra. Due to limitations in spectral fitting below 10 keV, it is necessary to additionally fix the spectral index and electron temperature in \texttt{Thcomp} model to their best-fitting values obtained from spectral fitting of joint NICER and NuSTAR spectra. In Figure~\ref{fig:5}, we present the plots of $F_{\rm disk}$ as a function of $T_{\rm in}$, $f_{\rm c}$, $N_{\rm H}$ and $p$ based on the fitting results obtained from the modelings of all 150 phase-resloved spectra. Regarding the relationship between $T_{\rm in}$ and $F_{\rm disk}$ (Figure~\ref{fig:5} (a)), we conduct fittings using the model $F_{\rm disk}\propto T_{\rm in}^\alpha$, resulting in $\alpha=3.03 \pm 0.27$ and $\alpha=5.07 \pm 1.21$ for spectral analyses with Models 1 and 2, respectively. It is worth noting that the derived value of $\alpha$ from spectral analysis with Model 1 are consistent with that presented in Figure~\ref{fig:4}. In addition, a weak anti-correlation between $F_{\rm disk}$ and $f_{\rm c}$ is observed from both fitting with Models 1 and 2. In the case of Model 1, $N_{\rm H}$ is treated as a free parameter, revealing a positive correlation with $F_{\rm disk}$ (see Figure~\ref{fig:5} (c)). Furthermore, when fitting these spectra with Model 2, a strong correlation between the $p$-value and $F_{\rm disk}$ is identified. Figure~\ref{fig:5} also demonstrates that the results of phase-resolved spectral fitting from all NICER observations are consistent with those obtained from the joint analysis of quasi-simultaneous observations from NICER and NuSTAR.

\section{Discussion} \label{sec:4}
In this study, we have conducted a detailed phase-resolved analysis of the newly identified Class \uppercase\expandafter{\romannumeral10} ``heartbeat"-like variability in IGR J17091--3624, utilizing data from NICER and NuSTAR observations. By employing the HHT method, we obtain phase-folded light curves in different energy bands, revealing a significant ($>10\sigma$) deficit/shortage in the hard X-ray ($>20$ keV) emission during the flare peak phases. Through modeling joint NICER and NuSTAR phase-resolved spectra, we observe strong variability in disk emission during the flare cycle, with synchronous variations in both disk temperature and flux with the count rate. In addition to the disk component, variations in the non-thermal component are also observed. Furthermore, spectral analysis of phase-resolved spectra from all NICER observations of the Class \uppercase\expandafter{\romannumeral10} state reveals robust correlations between spectral parameters and disk flux over the flare cycle.

Some other ``heartbeat"-like variability classes, such as well-known Class \uppercase\expandafter{\romannumeral4} in IGR J17091--3624 and Class $\rho$ in GRS 1915+105, also exhibit uniformity and coherence in flares. However, the flares of Class \uppercase\expandafter{\romannumeral10} exhibit greater symmetry compared to the typical Class $\rho$ variability \citep[show a \textit{slow rise and quick decay} feature,][]{2011ApJ...737...69N,2011ApJ...742L..17A,2017MNRAS.468.4748C,2024ApJ...963...14W}. In terms of the phase-resolved spectral analysis, spectral fitting with Model 1, we observe an increase in $\Gamma$ and a decrease in $kT_{\rm e}$ during the flare peak phases, similar to the behaviors reported by \citet{2011ApJ...737...69N} at the peak phases of the Class $\rho$ variability in GRS 1915+105. However, our findings reveal a subtle negative correlation between the covering fraction ($f_{\rm c}$) and disk flux, which contrasts with the previous study indicating that $f_{\rm c}$ reaches its maximum value near the peak phases of flares observed in the Class $\rho$ state. The key distinction between Class \uppercase\expandafter{\romannumeral10} and the Class $\rho$ variabilities may lie in the different evolution trends of the disk emission throughout the flare cycle. \citet{2011ApJ...737...69N} observed that GRS 1915+105 follows a clockwise trajectory in the $L_{\rm disk}$-$T_{\rm in}$ plot during the Class $\rho$ variability cycle. Specifically, following the trough phases, $L_{\rm disk}$ exhibits a steady increase at approximately constant disk temperature. Our findings show that the disk flux remains correlated with the temperature throughout the Class \uppercase\expandafter{\romannumeral10} variability cycle, albeit deviating from the standard $F_{\rm disk}\propto T_{\rm in}^4$ relationship. These comparisons between the Classes \uppercase\expandafter{\romannumeral10} and $\rho$ variability suggest that while both patterns of variability stem from instabilities within the accretion disk leading to significant fluctuations in disk emission, there exist some differences in their underlying physical processes.

\begin{figure}
\centering
    \includegraphics[width=\linewidth]{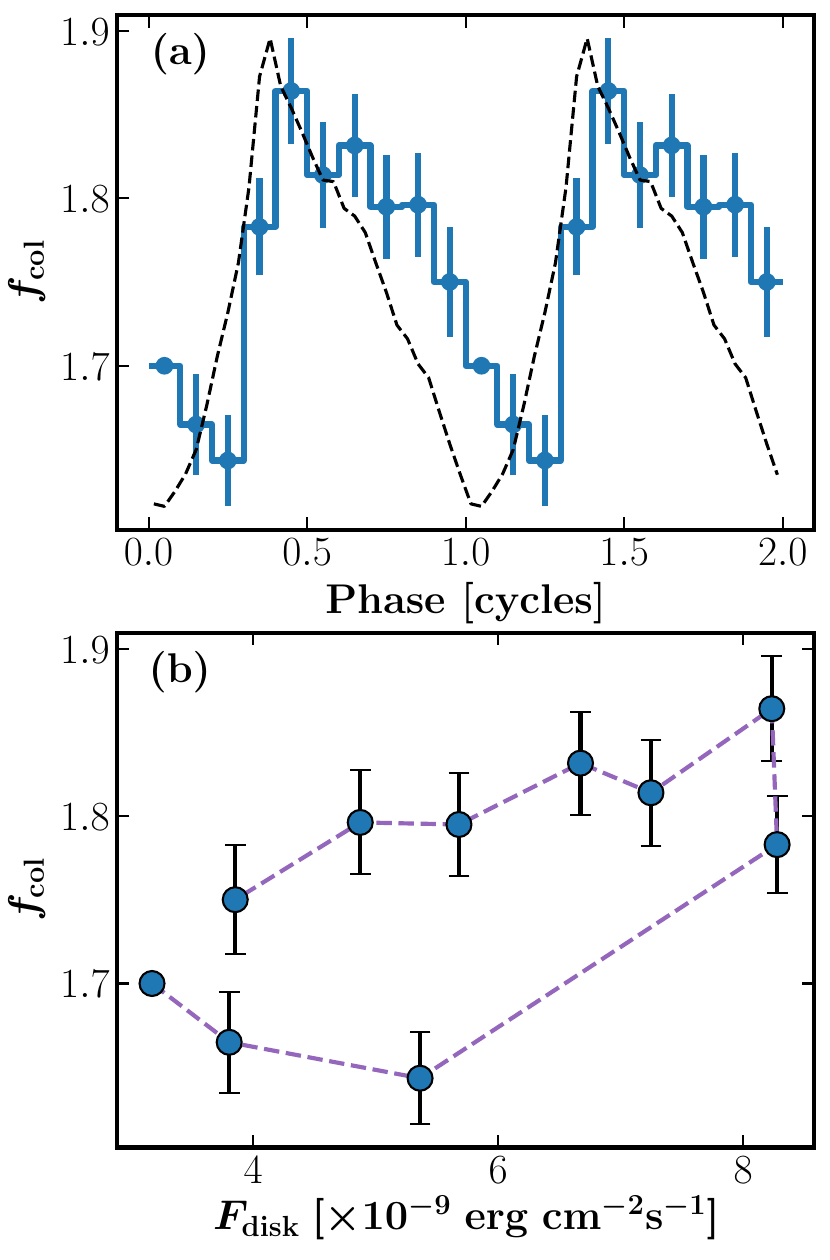}
    \caption{(a) Variations of hardening factor ($f_{\rm col}$), calculated using $f_{\rm col}=T_{\rm in}/T_{\rm eff}$. In this calculation, we assume that $f_{\rm col}=1.7$ at the trough phases (0--0.1 cycles) and $F_{\rm disk}\propto T_{\rm eff}^4$. (b) Plot of calculated $f_{\rm col}$ as a function of $F_{\rm disk}$ over the variability cycle.} \label{fig:6}
\end{figure}

\begin{figure}
\centering
    \includegraphics[width=\linewidth]{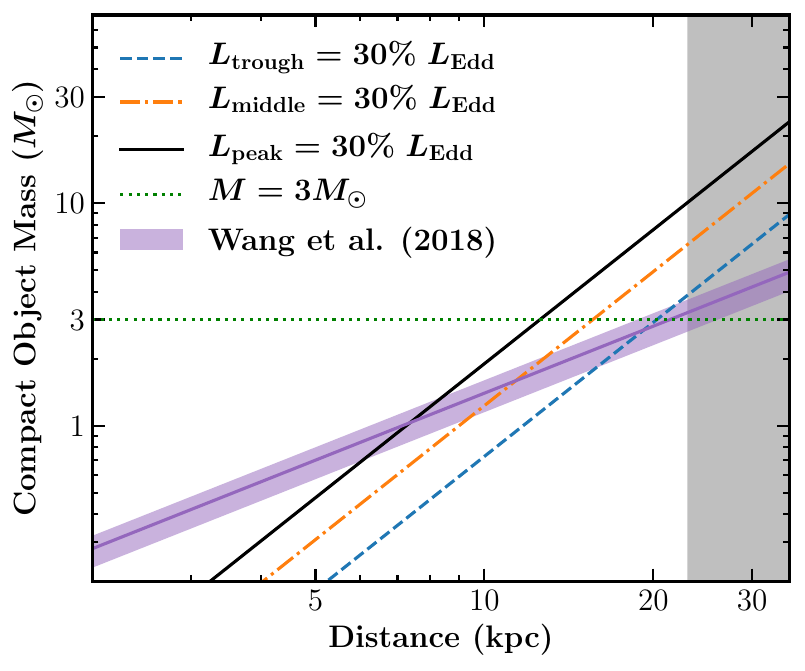}
    \caption{Plots compact object mass as a function of distance under different assumptions. The blue dashed line, orange dotted line and black solid line are obtained by assuming that the accretion disk emits at $30\%L_{\rm Edd}$ during the trough, middle and peak phases, respectively, of the Class \uppercase\expandafter{\romannumeral10} variability. The purple solid line with a confidence interval is taken from \citet{2018MNRAS.478.4837W}, which is plotted assuming that the inner radius deduced from the \texttt{Diskbb} model and reflection model are the same. The grey region indicates the distance of the source would be outside of the ``edge'' of the stellar disk in the galaxy \citep[see][]{2024ApJ...963...14W}.} \label{fig:7}
\end{figure}

Fitting the phase-resolved spectra using standard Model 1 yields a power-law index of $\alpha\sim3$ for the $L_{\rm disk}$-$T_{\rm in}$ correlation, which can be considered as an intermediate value between the standard-disk case of $\alpha=4$ and the slim-disk case of $\alpha=2$. The deviation of the disk emission from the expected $F_{\rm disk}\propto T_{\rm in}^4$ may stem from variations in the inner disk radius and hardening factor ($f_{\rm col}$), and departure of the disk temperature structure from the Shakura-Sunyaev prediction that $T(r)\propto r^{-3/4}$ \citep[see][]{2007A&ARv..15....1D}. If the hardening factor remains constant and the disk temperature has the radius dependence of $T(r)\propto r^{-3/4}$, our results indicate that the inner radius of the disk decreases from trough phases to peak phases, and then increases during the declining phases. On the other hand, if the inner radius is stably located at the innermost stable circular orbit (ISCO) during the flare cycle, the deviation in the $F_{\rm disk}$-$T_{\rm in}$ relation could also indicate a fluctuation in $f_{\rm col}$. Assuming that $f_{\rm col}=1.7$ \citep{1995ApJ...445..780S,2005ApJ...621..372D,2006ApJ...636L.113S} at the trough phases (0--0.1 cycles) and $F_{\rm disk}\propto T_{\rm eff}^4$, where $T_{\rm eff}$ is the effective disk temperature, we can calculate $f_{\rm col}$ for each phase bin using the formula $f_{\rm col}=T_{\rm in}/T_{\rm eff}$, where $T_{\rm in}$ is obtained from spectral fittings. Figure~\ref{fig:6} (a) shows variations in $f_{\rm col}$ over the flare cycle. We observe a slight phase difference between the disk flux and $f_{\rm col}$, but overall, there is a positive correlation between the two parameters (see Figure~\ref{fig:6} (b)). This finding is consistent with the prediction of disk models that the hardening factor increases with the disk flux at high luminosities \citep{2006ApJ...647..525D,2007A&ARv..15....1D}. The increase in $f_{\rm col}$ can be attributed to complete ionization of metals beyond temperatures of 0.9--1 keV, resulting in very little true opacity. An alternative explanation for the departure from the $L_{\rm disk}$-$T_{\rm in}$ relation could be attributed to a deviation in the temperature profile of the accretion disk from the expected $T(r)\propto r^{-3/4}$ scenario. This possibility has been taken into account in our spectral fitting using Model 2, where we introduce the power-law index ($p$) as a free parameter. Our analysis reveals that employing Model 2 for the spectral fitting partially rectifies the $L_{\rm disk}$-$T_{\rm in}$ relation; however, the derived value of $p$ exhibit fluctuations between 0.6 and 0.9 throughout the entire flare cycle. Considering the possible impact of Compton scattering on the $p$ value \citep{1994ApJ...426..308M}, this observed modulation may also be indicative of variations in $f_{\rm col}$. If the modulation in $p$ value is indeed genuine, it could suggest that instability in the accretion disk predominantly occurs in the inner region rather than the entire disk \citep{2007A&ARv..15....1D}. In this case, variations in accretion disk temperature also depend on the radius, resulting in higher apparent $p$ values at flare peak phases and lower values at trough phases.

It is widely accepted that the ``heartbeat"-like variability observed in GRS 1915+105 and IGR J17091--3624 is due to the radiation pressure instability which is known to occur only at high luminosities \citep{1998ApJ...508L..85V,2000A&A...355..271B,2000ApJ...535..798N,2002ApJ...576..908J,2007A&ARv..15....1D,2011ApJ...737...69N}. Meanwhile, our results reveal deviations of the disk emission from the expected $F_{\rm disk}\propto T_{\rm in}^4$ during the flare cycle, suggesting potential luminosity of $\gtrsim30\% L_{\rm Edd}$ \citep{2010A&A...521A..15A} for IGR J17091--3624 in the Class \uppercase\expandafter{\romannumeral10} state. Assuming an inclination of $46^{\circ}$ for the accretion disk of IGR J17091--3624, as obtained from the modeling of the reflection component \citep{2018MNRAS.478.4837W}, and emitting at $30\% L_{\rm Edd}$, we can derive the mass of the compact object as a function of distance for the fluxes at the trough ($\sim3.2\times10^{-9}\ \rm{erg\ cm^{-2}s^{-1}}$), middle ($\sim5.4\times10^{-9}\ \rm{erg\ cm^{-2}s^{-1}}$) and peak ($\sim8.3\times10^{-9}\ \rm{erg\ cm^{-2}s^{-1}}$) phases (see Figure~\ref{fig:7}), respectively. Additionally, assuming that the inner radius deduced from the \texttt{Diskbb} model and reflection model are the same, \citet{2018MNRAS.478.4837W} also presented a plot showing the mass of the black hole as a function of distance (plotted as purple line with a confidence interval in Figure~\ref{fig:7}). If the source emits at $30\% L_{\rm Edd}$ during the trough phases, the combination of the two analyses suggests a black hole mass of $\sim2-3M_{\odot}$ and a distance of $\sim17$--21 kpc. However, if the source emits at $30\% L_{\rm Edd}$ during the middle or peak phases, it could indicate that the estimated mass is $\lesssim2 M_{\odot}$, suggesting that the compact object of IGR J17091--3624 is more likely a neutron star.

In addition to the disk component, there is also a significant evolution in the hard X-ray emission throughout the Class \uppercase\expandafter{\romannumeral10} variability cycle. By phase folding the light curves, we detect shortages in the hard X-ray count rate within the energy range of $20$--79 keV at a confidence level of $\sim 15\sigma$ during the soft X-ray flare peak phases. The shortages in the hard X-ray of the continuum emission during type \uppercase\expandafter{\romannumeral1} bursts have been reported on several neutron star X-ray binaries \citep{2003A&A...399.1151M,2012ApJ...752L..34C,2018SSRv..214...15D}. One possible explanation for the observed shortages in hard X-ray emission is the scenario of corona cooling and recovery during type \uppercase\expandafter{\romannumeral1} bursts \citep[see e.g.][]{2012ApJ...752L..34C}. This scenario can also be employed to elucidate our findings: during the flare rising phase, the accretion disk undergoes a substantial rise in temperature and flux as a result of the radiation pressure instability, leading to the cooling of the corona by an abundance of seed photons. Subsequently, as the disk flux decreases, the corona recovers. The observed variability in electron temperature, which exhibits an anti-correlation with the modulations in disk temperature and flux, further supports this picture (see Figure~\ref{fig:3}). Studies of the shortages in the hard X-ray of the continuum emission during type \uppercase\expandafter{\romannumeral1} bursts generally found that the hard shortage lags the burst emission by $\sim1$ s, indicating that the fading and recovering of the hard X-ray flux follow the burst flux change almost instantaneously \citep[see e.g.][]{2012ApJ...752L..34C,2014A&A...564A..20J}. Through a cross-correlation analysis of the folded light curves using 50 phase bins within the energy ranges of 3--10 keV and 20--79 keV, we determine time lags of $0.25\pm0.32$ s and $0.07\pm31$ s for NuSTAR/FPMA and FPMB, respectively (see Appendix~\ref{appendix2}). This finding suggests a shorter timescale for corona heating in the Class \uppercase\expandafter{\romannumeral10} flare of IGR J17091--3624 compared to that observed in type \uppercase\expandafter{\romannumeral1} bursts of neutron star X-ray binaries, which is significantly shorter than the viscosity timescale. Furthermore, there is no significant phase lags observed between the disk flux and electron temperature.
It is challenging to explain a corona evolution on a second time-scale within the framework of traditional disk evaporation theory \citep{1994A&A...288..175M,1997ApJ...489..865E,2007ApJ...671..695L}, as these models attribute the formation of the corona to disk viscosity, implying a typical timescale of a few days. Alternatively, as discussed in \citet{2012ApJ...752L..34C} and \citet{2013MNRAS.432.2773J}, microscopic processes in the accretion disk, such as magnetic re-connection \citep{2000Sci...287.1239Z,2007HiA....14...41Z}, could potentially be self-consistently responsible for the rapid formation of the corona. However, current models do not indicate a characteristic magnetic field essential for re-connection within the disk, leaving the magnetic field of the accretion disk as an unknown factor in this process.
In addition to the variability in electron temperature, we also observe modulations in the covering fraction and spectral index throughout the flare cycle. The variations in the covering fraction may suggest that there are changes in the geometry of the corona during flares. For known values of spectral index and the electron temperature in the plasma, the Thomson optical depth ($\tau$) of the corona can be estimated using the theoretical relation \citep{1983ASPRv...2..189P,1987ApJ...319..643L,1993ApJ...413..507H}:
\begin{equation}
    \Gamma = -\frac{1}{2} + \sqrt{\frac{9}{4}+\frac{m_{\rm e}c^2}{kT_{\rm e}}\frac{1}{\tau(1+\tau/3)}},
\end{equation}
where $m_{\rm e}c^2 = 511$ keV is the rest mass energy of electron. If the optical depth remains constant, one would anticipate an anti-correlation between $\Gamma$ and $kT_{\rm e}$. However, we observe concurrent decreases in both $\Gamma$ and $kT_{\rm e}$ during the rising phases of the flares, indicating an increase in optical depth during these phases. In a scenario where a spherical corona partially covers the accretion disk, these variations in the Comptonization component may suggest that the corona undergoes effective cooling and contraction during the rising phase of the flare, and ultimately reaching its coolest and most compact state at the peak phases of the flare. However, during phases of $\sim0.5$--0.6 cycles, there is a notable sharp increase in $\Gamma$, accompanied by a corresponding rise in $kT_{\rm e}$, indicating a significant decrease in $\tau$ from $\sim6$ to $\sim1.9$ on a short time scale of $\sim6$ s. We propose that, during phases of $\sim0.5$--0.6 cycles, the material in the corona may be pushed away from the compact star by the radiation pressure of the disk emission, resulting in a sudden expansion of the corona region. This expansion could lead to a significant decrease in optical depth, despite an increase in the size of the corona. Following this expansion, as the luminosity of the disk decreases, the corona gradually contracts and the optical depth increases. %Additionally, the radiation pressure from the accretion disk may also produce outflows that could possibly induce changes in the circumstellar environment, contributing to observed fluctuations in absorption column density. 
However, the underlying physical mechanisms behind the proposed scenario regarding the expansion of the corona under the radiation pressure of the accretion disk lack clear definition and require further detailed analyses to assess its feasibility.

In Figure~\ref{fig:3}, we also observe the modulations in the absorption column density, which cannot be well studied in GRS 1915+105 due to its high foreground absorption. Similar modulations in $N_{\rm H}$ have been reported by \citet{2012ApJ...757L..12R}. However, we also find this modulation might be somewhat dependent on specific modeling choices, as satisfactory goodness of fit can still be achieved even when fixing $N_{\rm H}$ in the phase-resolved spectral fitting with Model 2. We propose that if the detected modulation in the absorption column density is indeed physical, it could be indicative of the disk wind launching process during the flare, particularly considering that \citet{2024ApJ...963...14W} have reported the detection of the absorption lines during this ``heartbeat"-like variability state in the 2022 outburst. According to \citet{2024ApJ...963...14W}, the absorption lines show blueshifts of up to 0.08 keV, indicating a very slow outflow velocity ($v_{\rm wind}<0.01c$). Assuming the outflow is triggered at the trough phase (i.e. 0 cycles), the outer radius of the outflow region at the peak phase of the flare (i.e. 0.4 cycles) is estimated to be $r_{\rm wind}\approx v_{\rm wind}\times t<7.8\times10^9 {\rm cm}$, where $t\approx26$s represents the duration of outflow propagation in the rising phase of the flare. Given the change in $N_{\rm H}$ from the trough to the peak phases to be $\sim2\times10^{21} \rm{cm^{-2}}$ and a spherical outflow region, the number density of the outflow ($n_{\rm wind}$) is inferred to be $\gtrsim2.6\times10^{11} {\rm cm^{-3}}$. Subsequently, the average mass outflow rate during the rising phase can be calculated as follows: $\dot{m}_{\rm wind}\approx\frac{4}{3}\pi r_{\rm wind}^3 m_{\rm p} n_{\rm wind}/t\lesssim3.3\times10^{16}\ \rm{g\ s^{-1}}$, where $m_{\rm p}$ is the proton mass. For comparison, the accretion luminosity can be expressed as $L_{\rm acc}=\eta \dot{m}_{\rm acc}c^2$, where $\eta$ is an efficiency factor typically taken to be 10\%, and $\dot{m}_{\rm acc}$ is the accretion rate. Assuming a source with a $3M_{\odot}$ black hole and mass accretion at a 30\% Eddington limit, we have $\dot{m}_{\rm acc}\approx1.3\times10^{18}\ \rm{g\ s^{-1}}$, indicating $\dot{m}_{\rm wind}/\dot{m}_{\rm acc}\lesssim0.03$. This measured outflow rate is significantly lower than that derived from the modeling of absorption lines during the 2011 outburst by \citet{2012ApJ...746L..20K}, consistent with the finding that the measured outflow velocity during the ``heartbeat" state in the 2022 outburst ($<0.01c$) is substantially reduced compared to that observed in the 2011 outburst \citep[$\approx0.05c$, see][]{2012ApJ...746L..20K}. It is worth noting that the aforementioned measurement of the outflow rate is qualitative, and a more accurate calculation requires precise spectroscopic analysis of the absorption line, which is beyond the scope of this work.

\section{Summary} \label{sec:5}
We have conducted a comprehensive phase-resolved analysis of the recently discovered ``heartbeat"-like variability Class \uppercase\expandafter{\romannumeral10} in IGR J17091--3624, during its 2022 outburst, utilizing data from NICER and NuSTAR observations. Differing from the classical ``heartbeat" variability, Classes $\rho$ in GRS 1915+105 and \uppercase\expandafter{\romannumeral4} in IGR J17091--3624, Class \uppercase\expandafter{\romannumeral10} variability shows a higher degree of uniformity and symmetry in its flares. By phase folding the light curves at high energies ($>20$ keV), a shortage in hard X-ray flux is observed at peak phases of the flare. Furthermore, a phase-resolved spectral analysis presents variability in the spectral shape, particularly revealing significant synchronous variations in the disk temperature and flux with the count rate. These findings suggest that the flare is primarily driven by some instabilities in the accretion disk, consistent with previous studies on Class $\rho$ variability. However, we also observe a positive correlation between the disk temperature and flux over the flare cycle, which differs from a loop relation between the two parameters found in the Class $\rho$ variability. This suggests potential differences in underlying physical processes between Classes $\rho$ and \uppercase\expandafter{\romannumeral10} variabilities. The variations in the Componization component during flares are also observed, with the electron temperature and covering fraction showing anti-correlations with the disk flux, revealing potential interactions between the accretion disk and the corona during the flare. These findings offer a comprehensive view of spectral behaviors of the newly identified ``heartbeat"-like variability, and reveal possible interaction processes between the accretion disk and the corona. Future phase-resolved analyses of the polarization data from IXPE and eXTP missions could offer further valuable insights into the nature of the ``heartbeat"-like variability.

\begin{acknowledgments}
We are grateful to the anonymous referee for constructive comments that helped us improve this paper. This research has made use of data obtained from the High Energy Astrophysics Science Archive Research Center (HEASARC), provided by NASA’s Goddard Space Flight Center. This work is supported by the National Key R\&D Program of China (2021YFA0718500) and the National Natural Science Foundation of China under grants, 12333007, 12173103, 12027803, U2038101 and U1938103. This work is partially supported by International Partnership Program of Chinese Academy of Sciences (Grant No.113111KYSB20190020). L. D. Kong is grateful for the financial support provided by the Sino-German (CSC-DAAD) Postdoc Scholarship Program (57251553). P. J. Wang is grateful for the financial support provided by the Sino-German (CSC-DAAD) Postdoc Scholarship Program (57678375).
\end{acknowledgments}





\appendix
\section{MCMC Parameter Probability Distributions}
\label{appendix1}
\begin{figure*}
\begin{minipage}[c]{0.48\linewidth}
\centering
    \includegraphics[width=\linewidth]{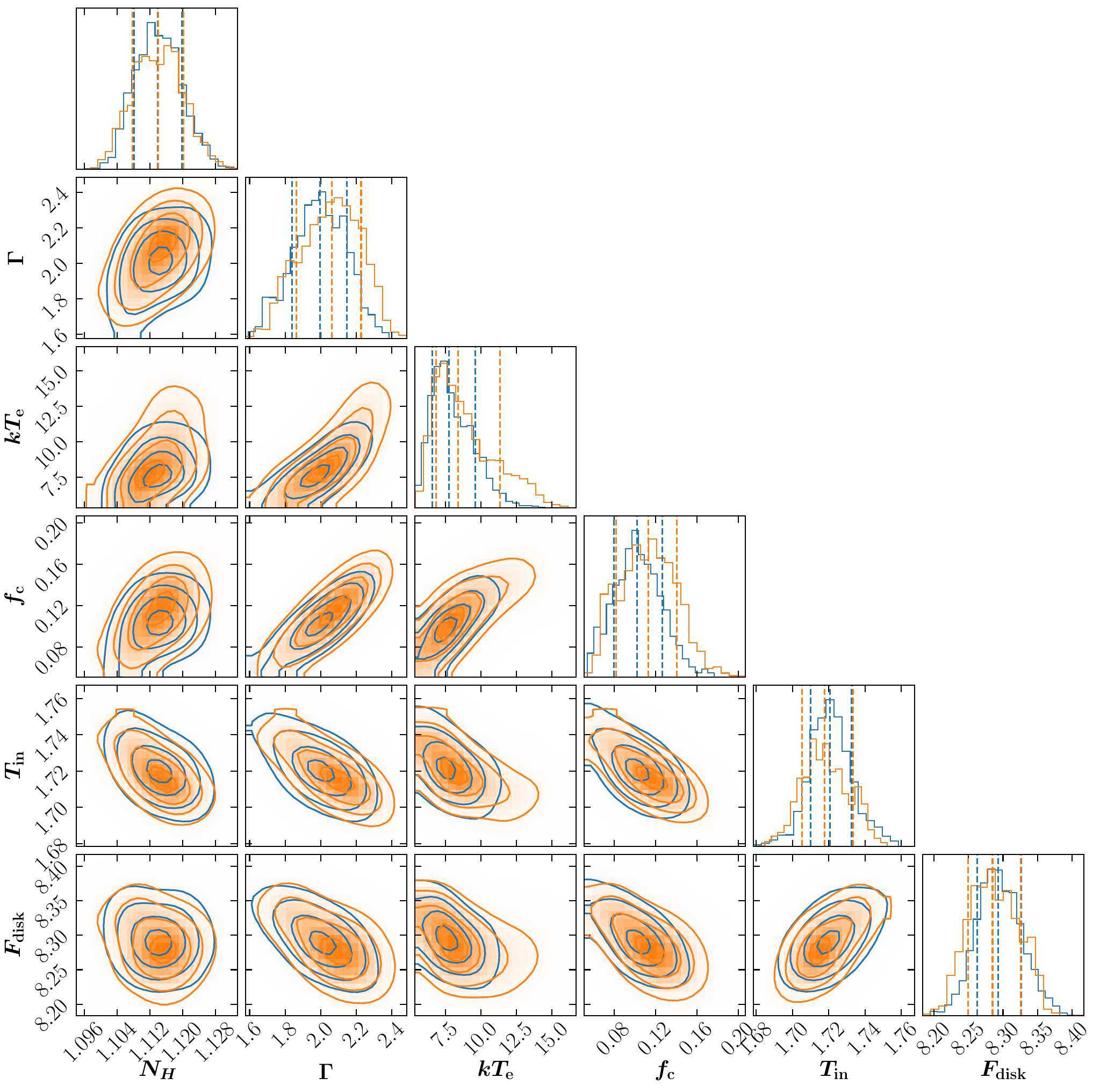}
\end{minipage}
\begin{minipage}[c]{0.48\linewidth}
\centering
    \includegraphics[width=\linewidth]{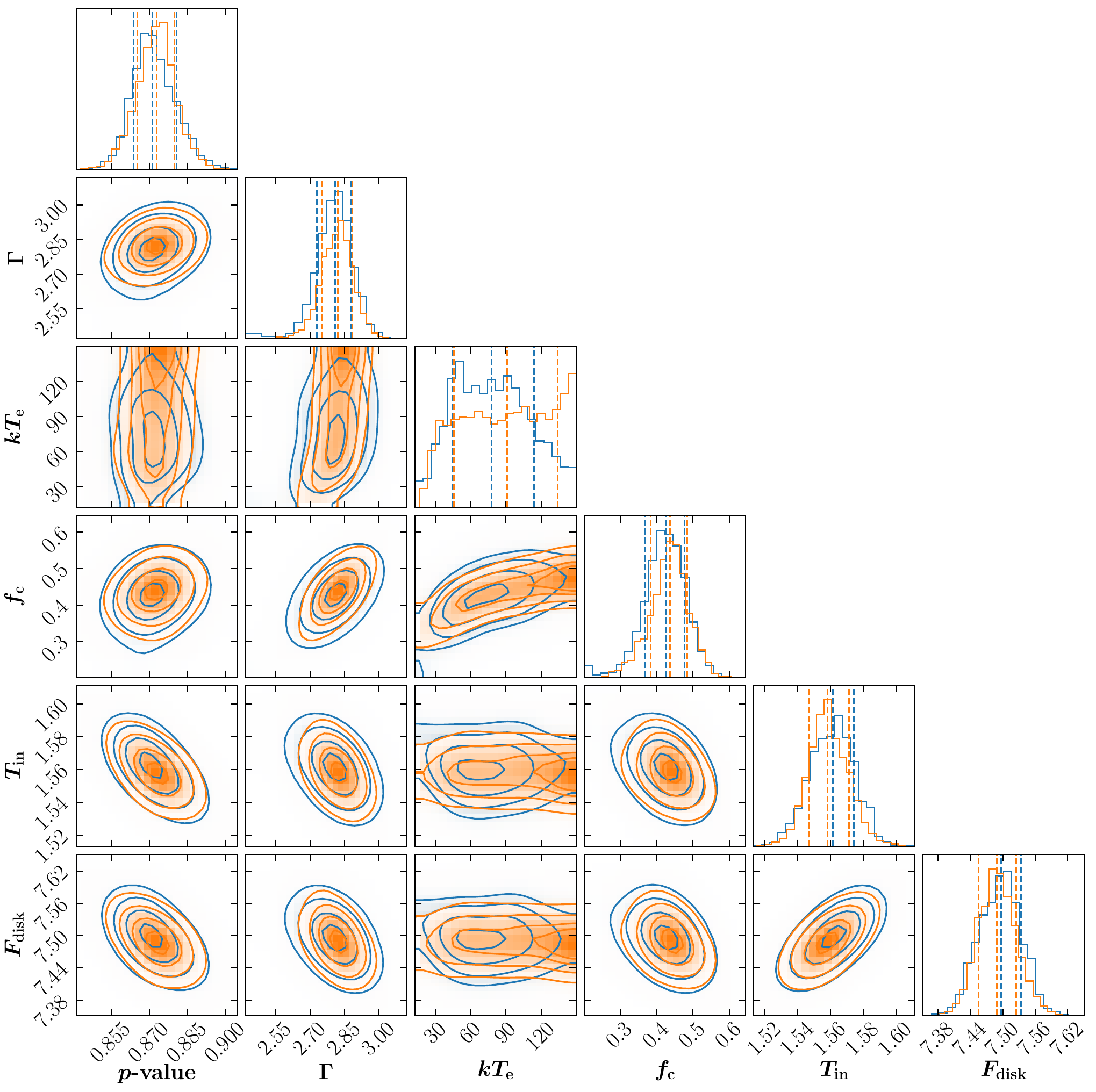}
\end{minipage}\\
\begin{minipage}[c]{0.48\linewidth}
\centering
    \includegraphics[width=\linewidth]{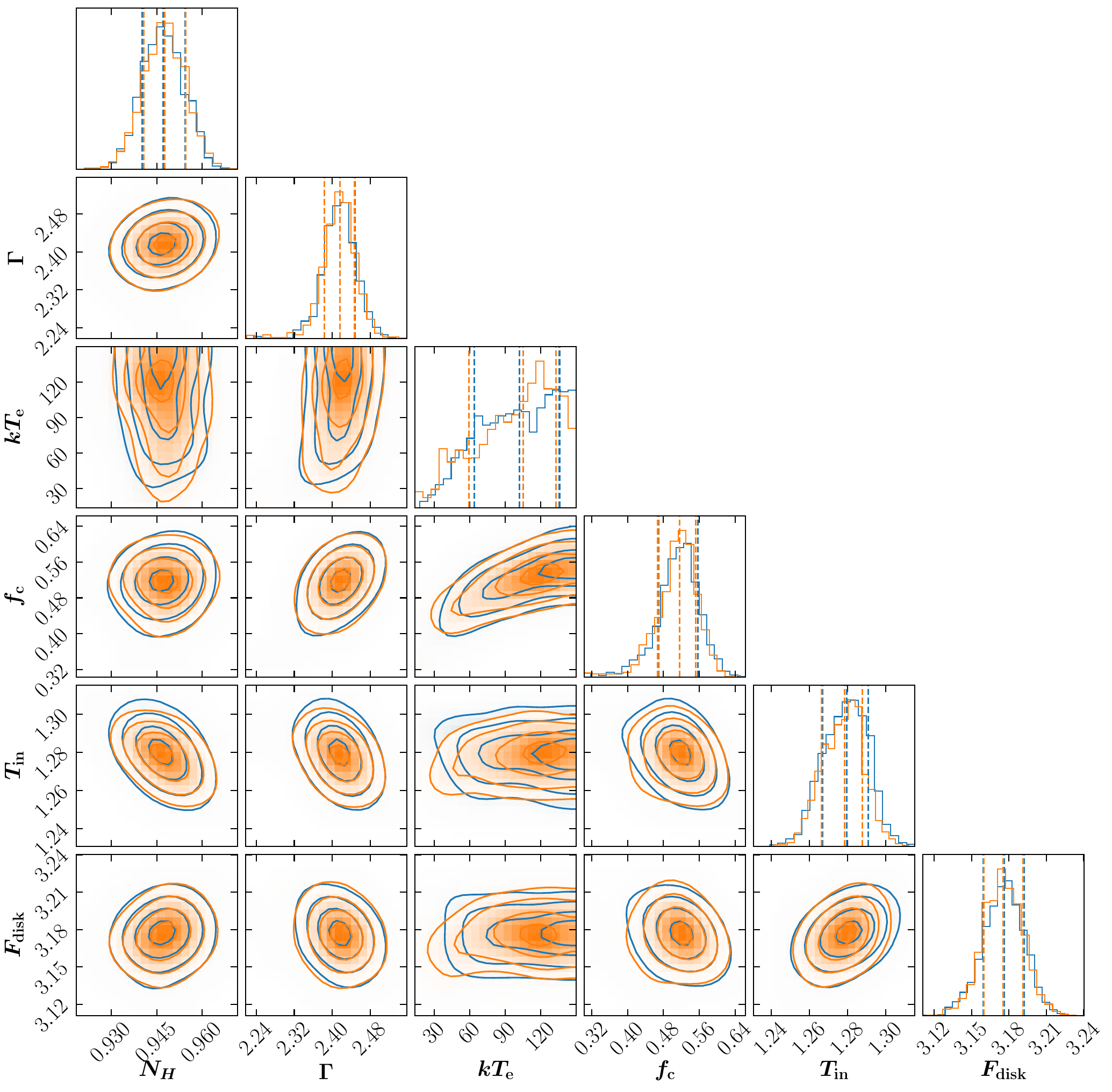}
\end{minipage}
\begin{minipage}[c]{0.48\linewidth}
\centering
    \includegraphics[width=\linewidth]{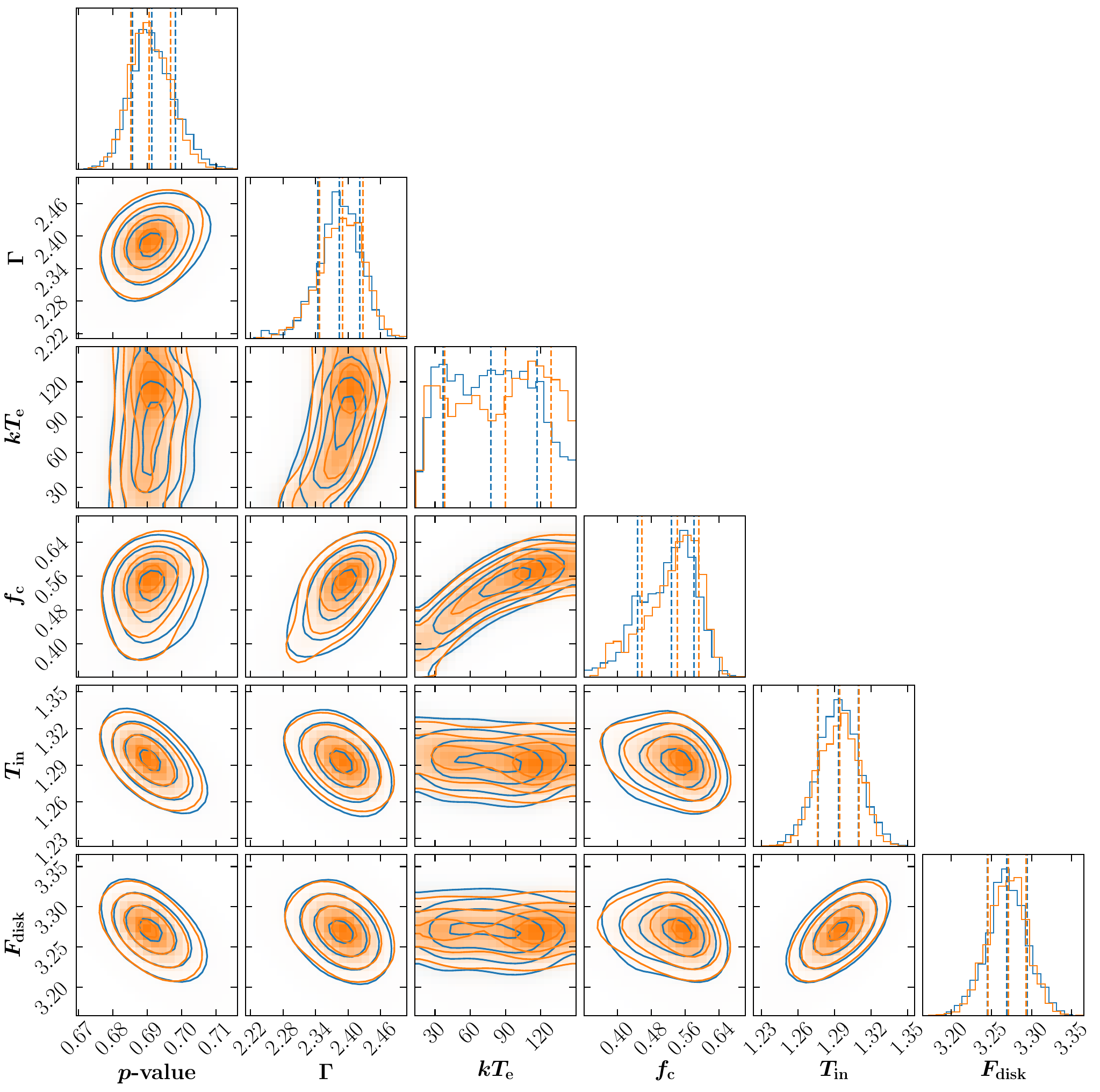}
\end{minipage}
    \caption{One- and two-dimensional projections of the posterior probability distributions, and the 0.16, 0.5, and 0.84 quantile contours derived from the MCMC analysis for each free spectral parameter, from the fitting of the joint NICER and NuSTAR phase-resolved spectra. To test the convergence, we compare the one- and two-dimensional projections of the posterior distributions from the first (blue) and second (orange) halves of the chain. These illustrations corresponds to the spectral fitting of the flare peak phases (top) and trough phases (bottom) with Model 1 (left) and Model 2 (right).} \label{fig:A}
\end{figure*}

This appendix contains corner plots of spectral parameters from an example MCMC analysis. We use the Goodman–Weare algorithm with 32 walkers and a total length of 50,000 to perform the MCMC analysis, and the initial 2000 elements are discarded as the burn-in period during which the chain reaches its stationary state. In Figure~\ref{fig:A}, we compare the one- and two-dimensional projections of the posterior distributions for each parameter from the first and second halves of the chain to test the convergence. The contour maps and probability distributions are plotted using the corner package \citep{2016JOSS....1...24F}. The electron temperature is effectively constrained at the peak phases of the flare using Model 1, while spectral fitting with Model 2 does not yield effective constraints. However, neither of the two modelings can effectively constrain the electron temperature at the flare trough phases.

\section{Determination of time lags}
\label{appendix2}
\begin{figure}
\centering
    \includegraphics[width=0.5\textwidth]{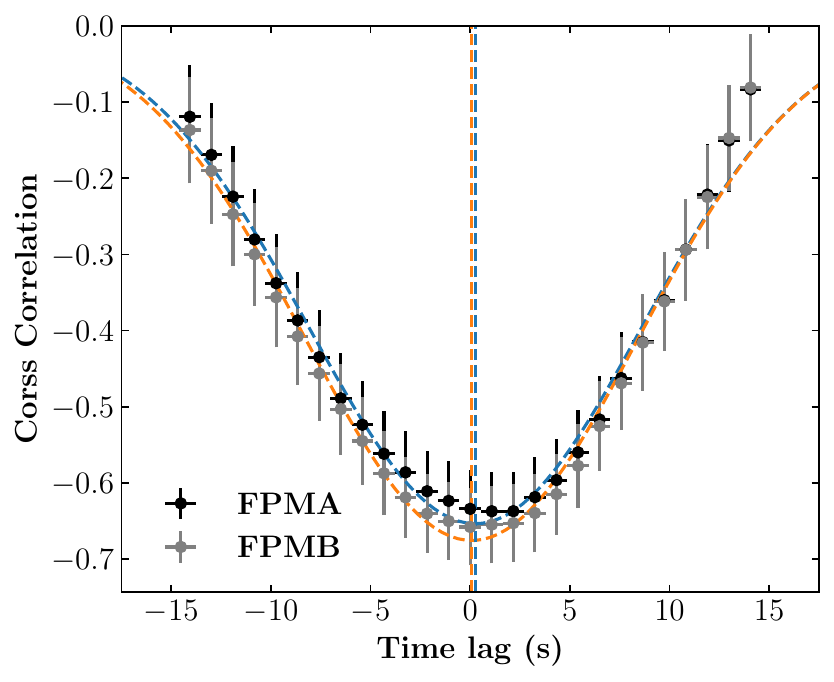}
    \caption{The cross-correlation between phase-fold light curves of 3--10 keV and 20--79 keV for FPMA and FPMB, respectively. The phase-folded light curves are derived in 50 phase bins. The curves show Gaussian fits to the data for estimating the time lags in these two energy bands, and the vertical dashed lines represent the determined time lags. The y-axis denotes the cross-correlation factor derived between the phase-folded light curves in the 3--10 keV and 20--79 keV bands.} \label{fig:B}
\end{figure}

This appendix presents the determination of the time lag between the shortage in hard X-ray (20--79 keV) phase-folded light curve and the flare in soft X-ray (3--10 keV) one by a cross-correlation method. We generate phase-folded light curves with 50 phase bins for the high- and low-energy energy bands to calculate the cross-correlation between them (see Figure~\ref{fig:B}). The error bar of the cross-correlation is computed with Poisson sampling of the phase-folded light curves. Subsequently, a negative Gaussian function is employed to fit the cross-correlation function. The determined time lags are $0.25\pm0.32$ s and $0.07\pm0.31$ s for NuSTAR/FPMA and FPMB data, respectively, assuming the recurrence time of the flare is $\sim 65$ s \citep[obtained from the averaged PSD analysis of][]{2024ApJ...963...14W}.

\bibliography{sample631}{}
\bibliographystyle{aasjournal}



\end{document}